\documentclass[vecphys]{svmult}

\usepackage{graphicx}       
\usepackage[bottom]{footmisc}

\newcommand{\eV}{\,\mathrm{eV}}

\newcommand{\Mpc}{\,\mathrm{Mpc}}

\newcommand{\km}{\,\mathrm{km}}

\newcommand{\sr}{\,\mathrm{sr}}
\newcommand{\yr}{\,\mathrm{yr}}

\begin{document}

\title*{Ultra High Energy Cosmic Rays: Observations and Theoretical Aspects}
\titlerunning{UHECRs: Observations and Theoretical Aspects}
\author{Daniel De Marco}
\institute{Bartol Research Institute, University of Delaware,
Newark, DE 19716, U.S.A.
\texttt{ddm@bartol.udel.edu}}

\maketitle

\begin{abstract}
We present a brief introduction to the physics of Ultra High Energy Cosmic Rays (UHECRs),
concentrating on the experimental results obtained so far and on what, from these results, can be
inferred about the sources of UHECRs.
\end{abstract}

\section{Introduction}
Since the discovery of cosmic rays (CRs) by Victor Hess in 1912 there has been a constant search for
the end of the cosmic-ray spectrum. This end has long been thought to be determined by the highest
energy the cosmic accelerators might be able to achieve, but despite several decades of research no
end of the spectrum was in sight until 1966. In 1966, right after the discovery of the cosmic
microwave background (CMB), it was understood~\cite{GZK} that protons with sufficiently high energy
would interact inelastically with the photons of the CMB and produce pions. This process rapidly
degrades the proton energy and, if the sources of CRs are homogeneously distributed, it produces a
drastic suppression in the CR flux around $10^{20}\eV$, where the so-called photo-pion production
starts to be kinematically allowed. The physical reason for this suppression is that at energies
around $10^{19}\eV$ the loss length for protons propagating in the CMB is of the order of a Gpc and
we are receiving particles from almost all the visible universe, whereas at $10^{20}\eV$ the loss
length is about $100\Mpc$ and we are receiving particles only from a tiny fraction of the universe.
This suppression is usually called the \emph{GZK cutoff}~\cite{GZK}. 
After several more decades of experimental activity we have now experiments exploring the energy
region around $10^{20}\eV$ and beyond, but the end of the spectrum is still eluding us and whether
the GZK cutoff is present or not in the observed spectra is still an open question.

In \S 2 we briefly review the present UHECR data sets and the issues they raise and in \S 3 we
briefly discuss how we can improve our understanding of the sources of UHECRs using the new data
that is now being collected by the Pierre Auger Observatory (PAO).

\section{UHECRs: present}
Until a year ago the two largest experiments measuring UHECRs were AGASA and HiRes. In 2005 the PAO
\cite{pao} reported the results of the first year of data taking \cite{augerdata}, but, since those
results are still preliminary and the error-bars are still quite large, in the following discussion
we will concentrate on AGASA and HiRes. We are considering particles of ultra high energy, around
and above $10^{19}\eV$, and CRs of such high energies, entering the earth atmosphere, interact with
it producing extensive showers of secondaries that propagate in the atmosphere close to the speed of
light. The two above mentioned experiments use two complementary techniques to detect these
extensive air showers (EAS): AGASA used an array of detectors on the ground that sampled the lateral
distribution of the EAS when it hit the ground while HiRes uses a telescope to observe the
fluorescence light produced by the shower while it propagates in the atmosphere. For a review of the
detection techniques see Ref.~\cite{nw}. 

Despite the fact that two completely different methods were used, the two experiments report somehow
similar results for what concerns the energy spectrum at low energy, with some conflicts at high
energy. At low energy, where the number of detected events per energy bin is big the two experiments
report fluxes that differ by about a factor 2, but this discrepancy can be accounted for by correcting
for the systematic errors on the energy determination reported by the two collaborations, about
$\pm15\%$. Doing this shift the two experiments agree perfectly in this energy range ($\le
10^{20}\eV$) \cite{DBO1,ddmts}.

At $E>10^{20}\eV$, where the statistics of events is very sparse due to the steepness of the spectrum of
CRs, the two experiments report opposite results: AGASA claims~\cite{agasaspec} to have observed a
continuation of the spectrum beyond the expected cutoff whereas HiRes claims~\cite{HiRespec} to have
observed the expected suppression. The statistics of events above $10^{20}\eV$ is however really
small and the discrepancy between the two experiments is just about $3\sigma$. Taking into account
the systematic errors as in the low energy region this discrepancy is reduced to about
$2\sigma$~\cite{DBO1}. Recently the AGASA collaboration revised down their energy assignments
\cite{teshimacris} by about 10\% further reducing the alleged discrepancy.

Even if nowadays the presence of the GZK suppression in the spectrum seems more plausible than its
absence the fact still remains that events with energies above $10^{20}\eV$ have been measured
several times by different experiments. Where did those particles come from? For astrophysical
accelerators it is extremely challenging to accelerate particles to such high energies \cite{hillas}
and in the few plausible models the sources are usually too far away from us for the particles to be
able to propagate to the earth without suffering sensible energy losses. Indeed no suitable sources
have been found within reasonable distances around the arrival directions of the highest energy
events. For a review on the origin of UHECRs see Ref.~\cite{blasirev2} and references therein. 

From the measurement of an EAS we can basically obtain three informations about the primary
particle: its energy, its direction and its nature or chemical composition. Each one of these
informations is important and can provide useful clues about the sources. Some experimental
techniques are better suited to measure one of them or another one \cite{nw}, but in general all the
experiments are in the end reconstructing those three quantities. We already discussed the energy
spectrum, we will skip the discussion of the chemical composition (for a review of the experimental
results see Ref.~\cite{chemicalcompo} while for discussions of the interesting problem of the
galactic--extra-galactic transition see Ref.~\cite{ABBGGH,allard}) and we will now concentrate on
the arrival directions of those UHE events.

AGASA reported \cite{agasaani} the presence of clustering in its set of events with energies above
$4\!\times\!10^{19}\eV$. While on large scales the arrival directions of these events appear to be
isotropic, on small scales they appear to arrive in clusters. AGASA observed 6 doublets and 1
triplet with angular separation less than $2.5^\circ$ on a set of about 70 events. These data point
in the direction of astrophysical point sources with a density of about $10^{-5}\Mpc^{-3}$, with
large error bars of about one order of magnitude. Combining this result with the energy spectrum it
is possible to obtain information about the luminosity of the sources themselves \cite{BD04,DBO2}.
The significance of this result is however still debated. First of all because the statistical
significance of the clustering signal, that in the beginning was quite high, turned out to be lower
in subsequent analyses \cite{fw} and also because HiRes did not see any anisotropy in its data set
\cite{fw2} though in this case too the statistical significance of the absence of clustering is not
very high \cite{sato}. Moreover it seems that the AGASA data set itself presents some internal
inconsistency and the probability of reproducing the AGASA result on the spectrum is reduced by a
large factor when taking into account a source density of $10^{-5}\Mpc^{-3}$~\cite{DBO1}.

\section{UHECRs: (near) future}
The PAO, being built in Argentina, is a new kind of experiment that combines the two above-mentioned
measuring techniques \cite{pao}. It consists of four fluorescence telescopes overlooking a ground
array of 1600 surface detectors covering an area of $3000\km^2$. The ground array exposure, above
$10^{19}\eV$, after 10 years of data-taking will be $70000\km^2\sr\yr$, to be compared for example
with $1645\km^2\sr\yr$ that was the AGASA exposure after 10 years of operation. About 10\% of the
detected events will be \emph{hybrid events}, detected at the same time by the ground array and by
the fluorescence telescopes. The Auger data set will help tremendously in our understanding of
UHECRs and of their sources. First of all because measuring hybrid events it will be possible to
solve the discrepancy in the energy assignments between fluorescence and surface detectors.
Moreover, with the huge statistics of events it will collect at high energy, the spectrum in the GZK
region will no longer be dominated by statistical fluctuations as in the AGASA and HiRes case and we
will be able to observe the presence or absence of the GZK feature in the spectrum and maybe the end
of the CR spectrum \cite{DBO1}. The huge statistics will be even more important to study the
anisotropies in the event arrival directions. The PAO will be able, already after a few years of
operations, to detect the presence of anisotropies both on large \cite{serpico} and small
\cite{DBO2} scales. For example it will be able to distinguish between a uniform distribution of
sources and a discrete distribution of sources with a given density already after 5 years if the
density is smaller than $10^{-3}\Mpc^{-3}$, whereas in order to distinguish between different
densities 15 years of operations are required or even a bigger experiment (for example Auger North).

{\bf Acknowledgments.}
This research is funded in part by NASA APT grant ATP03-0000-0080 at University of Delaware.

\end{document}